\documentclass[
]{ceurart}

\usepackage{listings}
\begin{document}

\copyrightyear{2026}
\copyrightclause{Copyright for this paper by its authors.
  Use permitted under Creative Commons License Attribution 4.0
  International (CC BY 4.0).}

\conference{The 1st Late Interaction Workshop (LIR) @ ECIR 2026,
  April 02, 2026, Delft, NL}

\title{Learn to Pool: Lightweight Fine-Tuning for Flexible Multi-Vector Compression}

\author[1]{Stefan Josef}[%
orcid=0009-0005-4387-6912,
email=stefan.m.josef@gmail.com,
]
\cormark[1]
\address[1]{Independent Researcher}

\cortext[1]{Corresponding author.}

\begin{abstract}
Late interaction models have shown strong generalization capabilities, often outperforming much larger dense embedding models. One challenge to their widespread deployment is the large number of token vectors they produce per document and the associated storage and memory costs. Pooling tokens at inference time has shown great promise to reduce the vector count with limited effects on retrieval accuracy. Large-scale pooling-aware training has demonstrated even more impressive results at high compression rates. We propose lightweight fine-tuning as a practical alternative and find that even minimal pooling-aware training with k-means yields broad gains over inference-only pooling, shows evidence of transfer across pooling methods and datasets, and --- with multi-factor training --- produces a single model effective across different compression levels. Our strongest model outperforms the unpooled baseline on BEIR SciFact across pool factors 1--6, implying a vector compression rate of 83\% at no cost to retrieval accuracy.
\end{abstract}


\maketitle

\section{Introduction}

Late interaction models such as ColBERT \cite{khattab2020colbert} represent documents as collections of token embeddings, allowing fine-grained token-level interactions that show strong generalization across domains and document lengths. A key challenge for their widespread adoption is the large number of vectors that need to be stored per document compared to single vector dense embedding models. 

Recent work has focused on reducing the number of vectors that need to be stored while retaining most of the accuracy of multi-vector models. Inference pooling as introduced by \citet{clavie2024pooling} applies hierarchical pooling to ColBERT models without any training and shows that the number of vectors can be reduced by $2$--$3\times$ without significantly lowering retrieval accuracy. \citet{veneroso2025crisp} has focused on improving the efficiency-accuracy curve by training pooling-aware late interaction models with k-means clustering. Models trained this way achieve impressive compression rates without performance degradation, yet they rely on large-scale contrastive training which may not be accessible or practical for many practitioners that care mostly about their in-domain datasets.

This paper investigates whether lightweight pooling-aware fine-tuning of existing ColBERT models can achieve gains over pure inference pooling techniques. We first evaluate a small and modern ColBERT model on selected datasets using training-free inference pooling techniques --- sequential (span), hierarchical or k-means clustering --- across various pool factors. We then fine-tune the model using different pool factors and methods. We evaluate all fine-tuned models against the untrained baseline model, assess cross-method spill-over and cross-dataset effects to capture potential degradation on other datasets. 

We find that:
\begin{enumerate}
 \item Hierarchical pooling is the strongest inference pooling technique overall, consistent with \citet{clavie2024pooling}.
 \item Lightweight fine-tuning is an effective technique for making models pooling-aware. K-means is the strongest training method overall, producing the most consistent gains over untrained baselines. However, the strength of the gains depends on pool factor and dataset. 
 \item Fine-tuning without pooling can severely degrade pooled performance, demonstrating that the observed gains are specifically attributable to including pooling in the training loop.
 \item Multi-factor fine-tuning (randomly sampling a different pool factor each batch) produces a single model that works well across different compression levels.
 \item Span-based pooling shows the largest relative gains from pooling-aware fine-tuning on NanoBEIR, but remains the weakest method overall.
 \item Cross-method spill-over: training with one pooling method can improve performance with other methods, suggesting that it makes the learned vectors more pooling-friendly overall.
 \item Cross-dataset transfer: pooling-aware fine-tuning with k-means largely improves pooled retrieval on other datasets, with only moderate impact on unpooled performance. 
\end{enumerate}

These findings suggest that practitioners can take an existing ColBERT model and, with a small investment in pooling-aware fine-tuning, achieve significant vector compression, with lower storage and memory costs, reduced scoring complexity, and retrieval accuracy that can match or even exceed the unpooled baseline at considerable compression levels.

\section{Related Work}

Pruning-based methods reduce the vector count by discarding tokens deemed less important. \citet{hofstaetter2022colberter} both aggregate subword tokens into whole-word representations and learn to remove stopwords, while \citet{liu2024matching} prune tokens post-hoc based on position, IDF, and attention scores.

\citet{clavie2024pooling} present a different approach, introducing token pooling. Rather than removing tokens of low importance entirely, pooling aggregates tokens by averaging their embeddings. Their method uses inference-only pooling, requiring no further training of a late interaction model. They evaluate three pooling methods --- sequential pooling, hierarchical clustering and k-means clustering --- and find that hierarchical clustering performs best. On a selection of BEIR datasets they find that hierarchical pooling improves over the unpooled baseline at pool factor 2 and still preserves 99.03\% relative performance on average at pool factor 3. This implies a 50\% vector reduction at no cost to retrieval accuracy and a 66\% reduction while accepting only a small deterioration. 

\citet{veneroso2025crisp} present CRISP, a detailed study of adding different pruning and pooling strategies into the training loop. They investigate fixed tail pruning, k-spacing as well as fixed and relative k-means pooling across query and document tokens. They trained all models on the substantial BGE dataset \cite{li2025bgedata} for 20k training steps. They find that fixed k-means pooling (always keeping the same number of query and document vectors) performs best. Their C8x32 model (keeps 8 pooled query and 32 document tokens) beats the unpooled baseline at 3.9x query and 2.9x document compression. They attribute this improvement to a denoising effect, suggesting that aggregating tokens into clusters can remove noise from the representations. Their C4x8 achieves an impressive 7.9x query and 11x document compression with only a 3.6\% loss in relative retrieval accuracy. 

\citet{yan2025docpruner} introduce DocPruner, which applies pruning to reduce patch embeddings in visual late interaction models. They compare different pruning- and pooling-based approaches in an inference-only setting. Their proposed method uses attention scores to estimate the importance of document patch embeddings and achieves 50--60\% compression rates with minor loss in retrieval accuracy. They find that for visual document retrieval, pruning-based approaches generally outperform methods based on pooling. However, they notice one important exception. Jina V4 \cite{guenther2025jinaembeddingsv4} is the only model where pooling-based methods approach (1d, 2d-pooling) or exceed (hierarchical clustering) their proposed DocPruner approach on ViDoRe-V2 \cite{mace2025vidorebenchmarkv2}. The authors suggest that this may result from the model's unique training style. \citet{guenther2025jinaembeddingsv4} train Jina V4 on an objective that has the model learn single and multi-vector representations simultaneously. This appears to make the learned vectors robust to inference pooling as a positive side effect. 

Motivated by the results of CRISP \cite{veneroso2025crisp} and the pooling resilience exhibited by Jina V4's multi-objective training \cite{guenther2025jinaembeddingsv4}, we investigate whether lightweight pooling-aware fine-tuning can improve upon the inference-only pooling methods introduced by \citet{clavie2024pooling}.

\section{Method}

\subsection{Pooling}
\label{subsec:method_pooling}

Token pooling is a simple technique to reduce the number of vectors per document in multi-vector models \cite{clavie2024pooling}. While pruning-based techniques remove token vectors with low importance entirely, pooling reduces the vector count by averaging multiple token embeddings. Since the definition of relevancy is dependent on a given query, which is not known at the time of document indexing, pooling offers a softer way of token compression than pruning. 

Following \citet{clavie2024pooling} we investigate three different pooling strategies:
\begin{enumerate}
 \item \textbf{Sequential (span) pooling:} We average-pool subsequent tokens in the document sequence. It produces approximately $\left\lfloor N/p \right\rfloor + 1$ pooled vectors (where $N$ is the number of document tokens and $p$ is the pool factor). For a pool factor $p=3$, span pooling simply averages the token embeddings of every 3 subsequent tokens without any overlap. The \texttt{[CLS]} special token is protected from pooling and preserved as a single token vector. If the sequence length is not evenly divisible by $p$, the remaining tokens form a smaller final group that is averaged as usual. Since there is no clustering overhead during training or inference, this is also the cheapest method. 
 \item \textbf{K-means pooling:} This technique uses standard k-means clustering \cite{MacQueen1967SomeMF} on L2-normalized token vectors to create $\left\lfloor N/p \right\rfloor + 1$ pooled vectors. After creating $\left\lfloor N/p \right\rfloor$ clusters, the token vectors within each cluster are averaged to create a single pooled cluster embedding. Finally, the \texttt{[CLS]} token is preserved as a single token vector as above. 
 \item \textbf{Hierarchical pooling:} This technique uses hierarchical clustering \cite{murtagh2012hierarchicalclustering} with Ward linkage \cite{ward1963hierarchical} to create at most $\left\lfloor N/p \right\rfloor + 1$ pooled vectors for each document. The token vectors within each cluster are then averaged to create a single pooled cluster embedding, while the \texttt{[CLS]} token is again preserved as a single token vector.
\end{enumerate}

Unlike span pooling, clustering-based methods do not guarantee that a pooled vector is composed from exactly $p$ token vectors, which will only be true on average. 
We apply pooling only to the document-side and leave queries unchanged, maintaining one vector per token. We leave additional query-side pooling, as explored by \citet{veneroso2025crisp} for future work. 
We evaluate pool factors 2, 3, 4, 5, and 6 for each technique, implying vector reductions of 50--83\%.

\subsection{Pooling-Aware Fine-tuning} 

To measure whether pooling-aware training with limited resources can have beneficial effects over inference-only pooling, we apply lightweight fine-tuning on a single dataset.

Each fine-tuning run uses the same base model \texttt{mxbai-edge-colbert-v0-32m} and settings: We use the fused AdamW optimizer, a learning rate of $10^{-5}$ with a linear schedule and 10\% warmup steps, and train for 5 epochs with batch size 16 and documents truncated to 300 tokens. We run NanoBEIR evaluation for the dataset we are fine-tuning on after every epoch and select the checkpoint with the highest NDCG@10. 

We fine-tune models:
\begin{enumerate}
 \item \textbf{with a fixed pool factor:} The model sees a single pool factor during training. We apply the same pool factor for each batch during training. We train models with pool factors 2 and 3. 
 \item \textbf{with multiple pool factors:} The model sees different pool factors during training. We randomly sample a different pool factor between 1 and 6 for each batch from a uniform distribution. 
 \item \textbf{without pooling:} To separate the effect of pooling-aware training from simply training on a dataset for longer, we also fine-tune the same model without any pooling for comparison. 
\end{enumerate}

\subsubsection{Training Data}

We select the training splits of two BEIR datasets for fine-tuning: SciFact (809 training queries) and FiQA (5,500 training queries). Each model is fine-tuned on a single dataset. For each dataset, we carry out the following data processing pipeline: For each query, we select positive documents from BEIR qrels and employ hard negative mining with \texttt{bge-small-en-v1.5}\footnote{\url{https://huggingface.co/BAAI/bge-small-en-v1.5}} to pick the top-100 hard negative candidates as well as 30 random negative candidates sampled from the full corpus. 

We then run \texttt{bge-reranker-v2-gemma}\footnote{\url{https://huggingface.co/BAAI/bge-reranker-v2-gemma}} to obtain reranker scores for each query-document pair. Based on the reranker scores, we select the top-1 positive and filter out all negative candidates with a score higher than 95\% of the positive score, following the approach presented in \citet{moreira2025nvretriever}. From the remaining negatives, we select the top-5 documents as hard negatives, sample another 5 from the remaining hard negative candidates, and 5 more from the pool of random negatives. This yields 15 negatives per query: a balanced mix of hard, medium, and easy negatives, while reducing the likelihood of including false negatives. 

\subsubsection{Distillation Loss}

Following the industry standard for training multi-vector models, we fine-tune all models with a distillation loss based on the teacher scores obtained from the reranker as described above. The key mechanism that makes training pooling-aware is applying pooling \emph{inside} the forward pass, before scoring and loss computation. For each training batch:
\begin{enumerate}
  \item All queries and documents are encoded by the ColBERT encoder and L2-normalised.
  \item A pool factor~$p$ is selected from the configured set~$\mathcal{P}$. For multi-factor training $\mathcal{P} = \{1,2,3,4,5,6\}$ and~$p$ is sampled uniformly per batch; for single-factor training the same fixed~$p$ is used for every batch. Document embeddings are then pooled with the configured method, reducing each document to approximately $\lfloor N/p \rfloor + 1$ vectors; when $p = 1$ the embeddings are left unchanged.
  \item ColBERT's MaxSim scoring computes the student score for each query--document pair on the \emph{pooled} representations:
    \[
      s_{i,j} \;=\; \sum_{k} \max_{l}\; \mathbf{q}_{i,k}^\top \hat{\mathbf{d}}_{i,j,l}
    \]
    where $\hat{\mathbf{d}}$ denotes the pooled document token embeddings and $j$ indexes the candidate documents for query $i$. Each query token~$k$ is matched to its most similar document token~$l$ via dot product, and the per-token maxima are summed to produce the final score. Student scores are then min-max normalised per query to align them with the teacher score scale, yielding $\bar{\mathbf{s}}_i$.
  \item The training loss is the KL-divergence between the softmax-normalised teacher (reranker) and student score distributions, without temperature scaling:
    \[
      \mathcal{L} \;=\; \frac{1}{B} \sum_{i=1}^{B} \mathrm{KL}\!\bigl(\,\mathrm{softmax}(\mathbf{t}_i) \;\|\; \mathrm{softmax}(\bar{\mathbf{s}}_i)\,\bigr)
    \]
\end{enumerate}

Span pooling is fully differentiable. For clustering-based methods, cluster assignments are computed from detached embeddings outside the computational graph. The resulting cluster-averaged representations, however, are computed on the original embeddings within the computational graph, enabling gradient flow back into the encoder. The model is therefore directly optimised to produce token representations whose MaxSim scores survive aggregation. For multi-factor training, the per-batch sampling of~$p$ forces the encoder to produce representations that work across all compression levels, including $p = 1$ (no pooling), which acts as an implicit regulariser preserving unpooled performance.

\subsection{Experimental Setup}

\begin{description}
  \item[Model.] We choose \texttt{mxbai-edge-colbert-v0-32m}\footnote{\url{https://huggingface.co/mixedbread-ai/mxbai-edge-colbert-v0-32m}} \cite{takehi2025fantasticsmallretrievers}, a modern, open, and performant 32-million-parameter ColBERT model for all experiments in this paper.

  \item[Frameworks.] We use well-established implementations of all training and evaluation techniques. All experiments were carried out in PyLate \cite{chaffin2025pylate} and Sentence Transformers\footnote{\url{https://www.sbert.net/}}. Hierarchical pooling uses the SciPy implementation, while k-means clustering uses scikit-learn.\footnote{We compared different implementations of k-means: fastkmeans on CPU and GPU and scikit-learn on CPU, and found that scikit-learn's implementation was both fastest and most accurate and is therefore used for all experiments in this paper.}

  \item[Indexing.] All NanoBEIR runs were evaluated using exact search with a fixed query length of 48 tokens and documents truncated to 300 tokens. BEIR evaluations used PLAID \cite{santhanam2022plaid} indexing via the fast-plaid backend with default parameters (4-bit product quantization, 8 IVF probes, 8,192 candidate documents for full reranking), top-20 retrieval depth, documents truncated to 300 tokens, and query lengths adjusted per dataset (32--64 tokens). All models were evaluated in fp16.

  \item[Evaluation Data.] To allow for a larger number of experiments while remaining within budget, we conducted most evaluations on selected NanoBEIR \cite{nanobeir2024} datasets: FiQA, NFCorpus, SCIDOCS, SciFact and Touché2020. We then assessed the main findings on the following BEIR \cite{thakur2021beir} datasets: FiQA, NFCorpus, SCIDOCS and SciFact. 

  \item[Metrics.] For all evaluation runs we report NDCG@10 as is standard in IR research. For each dataset we use the unpooled NDCG@10 score of the baseline model as the baseline, and report relative performance (in \%) of each pooling method and fine-tuned model.\footnote{Code for pooling-aware training and evaluation is available in our fork of PyLate: \url{https://github.com/stefan-jo/pylate}.}
\end{description}

\section{Results}

Unless otherwise noted, all results in Sections 4.1–4.2.4 are evaluated on NanoBEIR datasets. We then evaluate the main findings on full BEIR datasets in Section 4.2.5.

\subsection{Inference-Only Pooling}

We first evaluate the baseline model \texttt{mxbai-edge-colbert-v0-32m} on 5 NanoBEIR datasets using inference-only pooling. We compare sequential (span) pooling, hierarchical and k-means pooling as described in Section~\ref{subsec:method_pooling}. Consistent with the findings in \citet{clavie2024pooling}, hierarchical pooling leads to the best results overall, while span pooling performs worst overall. 

Hierarchical and k-means pooling both retain high relative performance at pool factor two. On average, hierarchical pooling achieves a relative score of 98.2\%. K-means achieves an average score of 96.1\% and span pooling only 89.9\%. Interestingly, at pool factor 3 k-means outperforms hierarchical pooling on average, driven mainly by very strong performance on NFCorpus. Hierarchical pooling stays above 90\% relative performance until pool factor 5, while k-means falls slightly below 90\% already at pool factor 4. Span pooling degrades much faster than clustering-based methods, retaining only 50\% relative performance at pool factor 5 on average. 

When looking at different datasets, it becomes obvious that some datasets are much more pooling-friendly than others. On Touché2020, hierarchical and k-means pooling stay (with one exception) above 90\% relative performance through pool factor 6, even slightly outperforming the unpooled baseline at pool factor 2. Other datasets are much less forgiving to inference-only pooling. On SciFact, both hierarchical and k-means pooling drop below 90\% relative performance at pool factor 4, further degrading below 80\% at pool factor 6. Span pooling degrades sharply on SciFact, retaining less than 30\% relative performance at pool factors 5 and 6.

When comparing these results to \citet{clavie2024pooling}, we notice steeper declines in relative performance across pool factors. As we will see in Section~\ref{subsubsec:beir_validation}, these differences are likely attributable to NanoBEIR, which seems overall less forgiving to inference pooling than corresponding BEIR datasets. We leave the investigation of these differences to future research.

\begin{figure}
  \centering
  \includegraphics[width=\linewidth, alt={Line chart showing relative NDCG at 10 declining as pool factor increases from 1 to 6 for hierarchical, k-means, and span pooling on five NanoBEIR datasets. Hierarchical pooling degrades most gracefully while span pooling drops steeply.}]{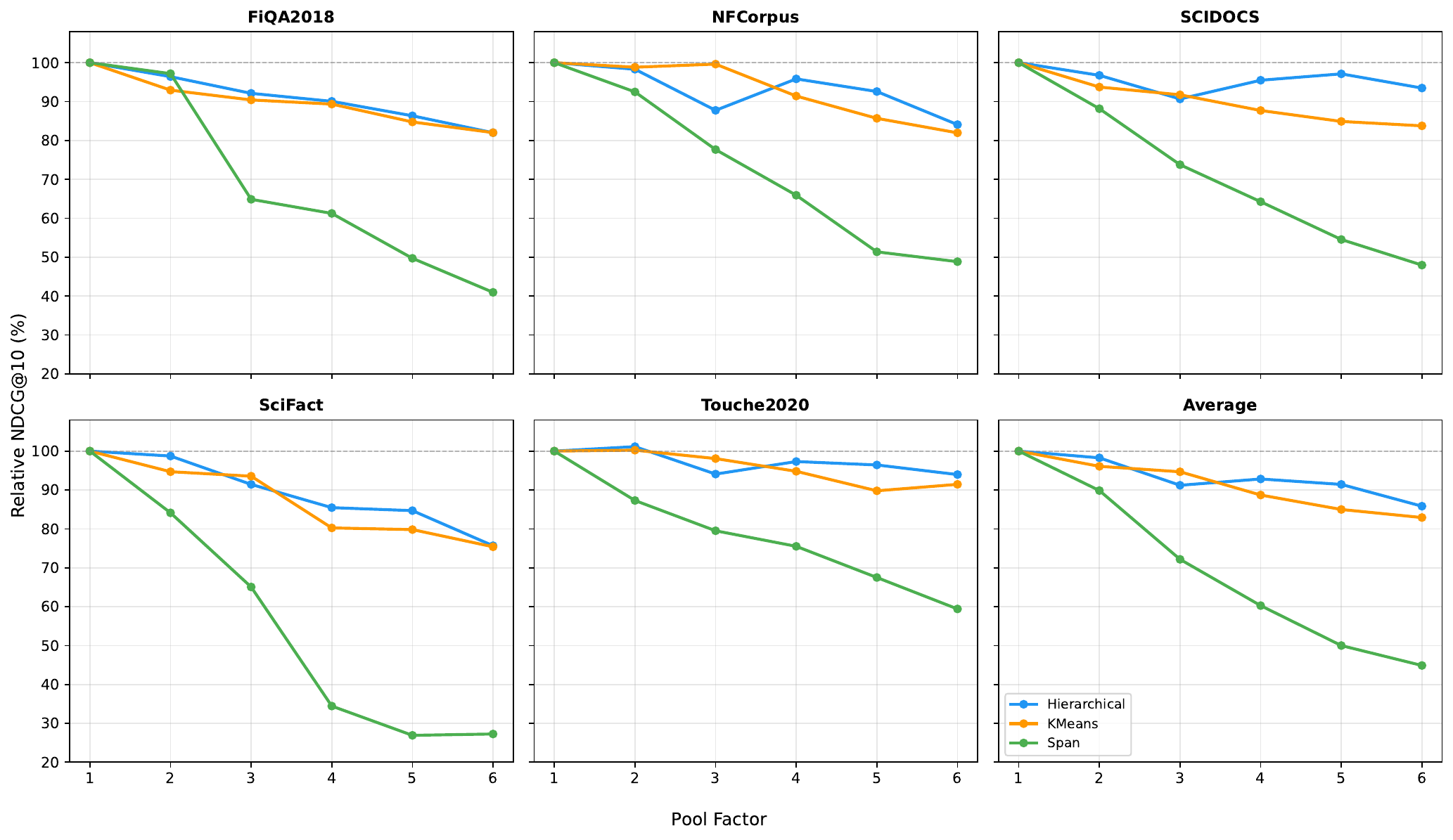}
  \caption{Relative NDCG@10 at various pool factors for three inference-only pooling methods applied to the baseline model \texttt{mxbai-edge-colbert-v0-32m} on NanoBEIR, without fine-tuning. }
  \label{fig:inference_pooling}
\end{figure}

\subsection{Pooling-Aware Fine-Tuning}

\subsubsection{Training-Aware Results}
\label{subsubsec:training_aware_results}

\begin{figure}
  \centering
  \includegraphics[width=\linewidth, alt={Two line charts comparing span pooling on SciFact and FiQA. Pooling-aware fine-tuned models retain much higher relative NDCG at 10 across pool factors 2 to 6 compared to the baseline, while fine-tuning without pooling severely degrades performance.}]{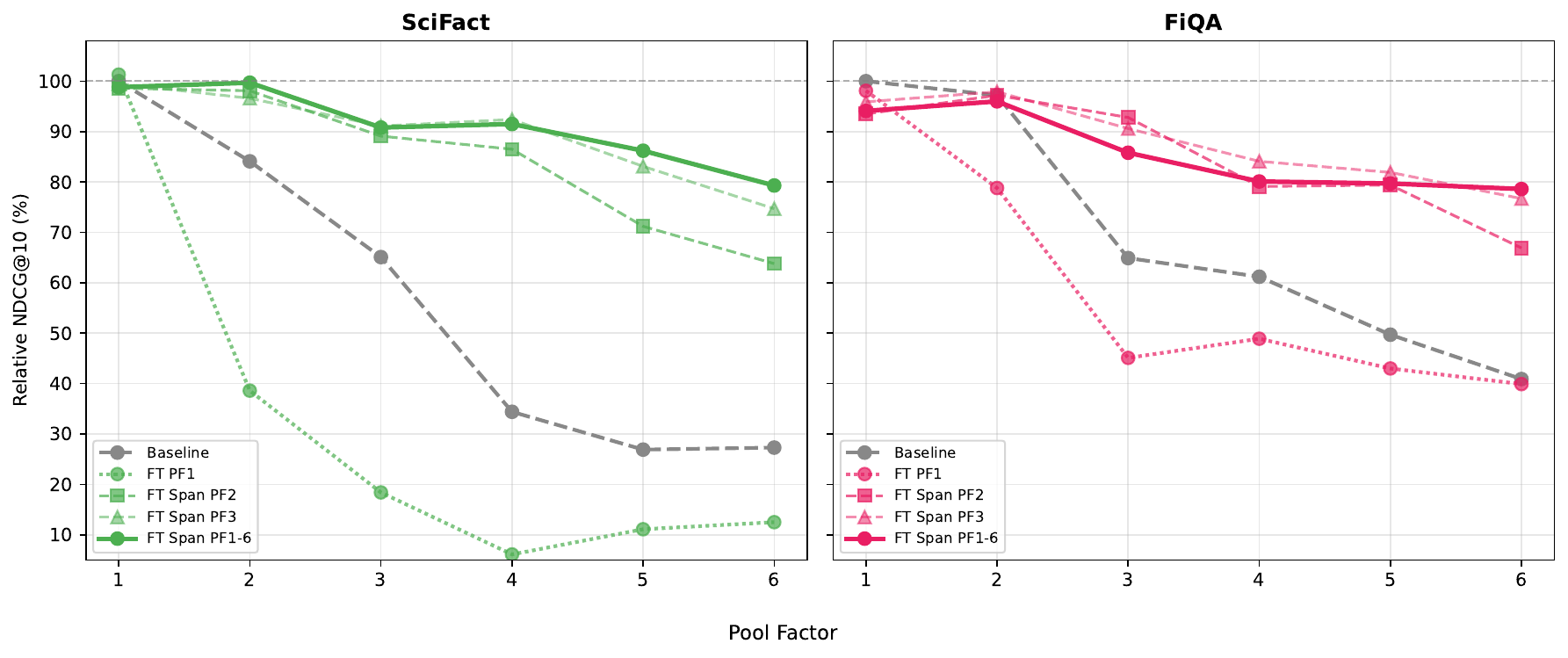}
  \caption{Effect of pooling-aware fine-tuning with span pooling on NanoBEIR SciFact (left) and FiQA (right). Each fine-tuning method is applied only to the dataset it is evaluated on. PF = pool factor. FT = fine-tuned; FT PF1 = fine-tuned without pooling during training; PF1-6 = multi-factor training with pool factors randomly sampled per batch.}
  \label{fig:span_scifact_fiqa}
\end{figure}

Figure~\ref{fig:span_scifact_fiqa} shows the effects of pooling-aware fine-tuning on two NanoBEIR datasets using span pooling. We observe that pooling-aware fine-tuning greatly recovers performance compared to inference-only pooling on both datasets. 

The effects are strongest on SciFact, where inference-only span pooling performed worst initially. Pooling-aware fine-tuning with a fixed pool factor of 2 (FT Span PF2) improves relative performance from 84.1\% to 98.1\% at pool factor 2. Gains become even stronger at higher pool factors. Training with a fixed pool factor of 3 (FT Span PF3) shifts the sweet spot of the curve, outperforming other methods on pool factors 3 and 4. Despite training on a single pool factor, we see gains across all pooled settings (PF$\,\geq$\,2), while accepting a small regression (98.6--99.5\%) on unpooled performance. 

Multi-factor training, where we randomly sample a pool factor between 1 and 6 at each batch, shows the best trade-offs overall: it actually performs best on pool factors 2 (retaining 99.7\% of performance), 5 (86.2\%) and 6 (79.3\%), while being a close second to FT Span PF3 for pool factors 3 and 4. This shows that we do not have to commit to a single fixed pool factor before training, allowing us to choose different pool factors at inference, with almost no cost compared to training on a single pool factor. 

On FiQA, we similarly observe clear gains from pooling-aware fine-tuning. Here, we find that training with a single pool factor of 3 shows the best performance overall, outperforming multi-factor through pool factors 1--5. While there is no large difference at pool factor 2, the model reaches 90.6\% relative performance at pool factor 3, compared to just 64.9\% for the baseline. In contrast to SciFact, we observe a stronger regression on unpooled performance on FiQA (93.5--95.9\%). 

Fine-tuning without pooling (FT PF1) actually hurts pooled performance compared to the baseline. On SciFact, unpooled performance slightly increases to 101.3\%, before collapsing to below 40\% at pool factor 2 and to 6.1\% at pool factor 4, far worse than the baseline (34.4\%). The effects on FiQA are less dramatic but still negative. This shows that naive fine-tuning has the potential to destroy a baseline model's inherent pooling compatibility and implies that the gains observed in pooling-aware training cannot be attributed to continued in-domain fine-tuning alone, but specifically to the inclusion of pooling in the training loop.

We hypothesize that this degradation results from the loss of regularization inherent in multi-task ColBERT pre-training: training across diverse datasets and tasks forces token representations to remain general, which may incidentally make them amenable to pooling. Fine-tuning on a single task without pooling allows the model to specialize its representations, losing this implicit pooling-friendliness.

Results for hierarchical and k-means pooling can be found in Appendix A, Table~\ref{tab:hier_training} and Table~\ref{tab:kmeans_training}, respectively. Overall we observe similar positive effects from pooling-aware fine-tuning as for span pooling: large gains over the baseline, unpooled fine-tuning destroys performance, fine-tuning on a fixed pool factor wins locally, while multi-factor training leads to the best results on average across all pool factors.

\subsubsection{Cross-Method Comparison}

\begin{figure}
  \centering
  \includegraphics[width=\linewidth, alt={Two line charts comparing multi-factor fine-tuned models using span, hierarchical, and k-means pooling on SciFact and FiQA. K-means fine-tuning achieves the strongest results, outperforming the unpooled baseline on SciFact across pool factors 1 to 4.}]{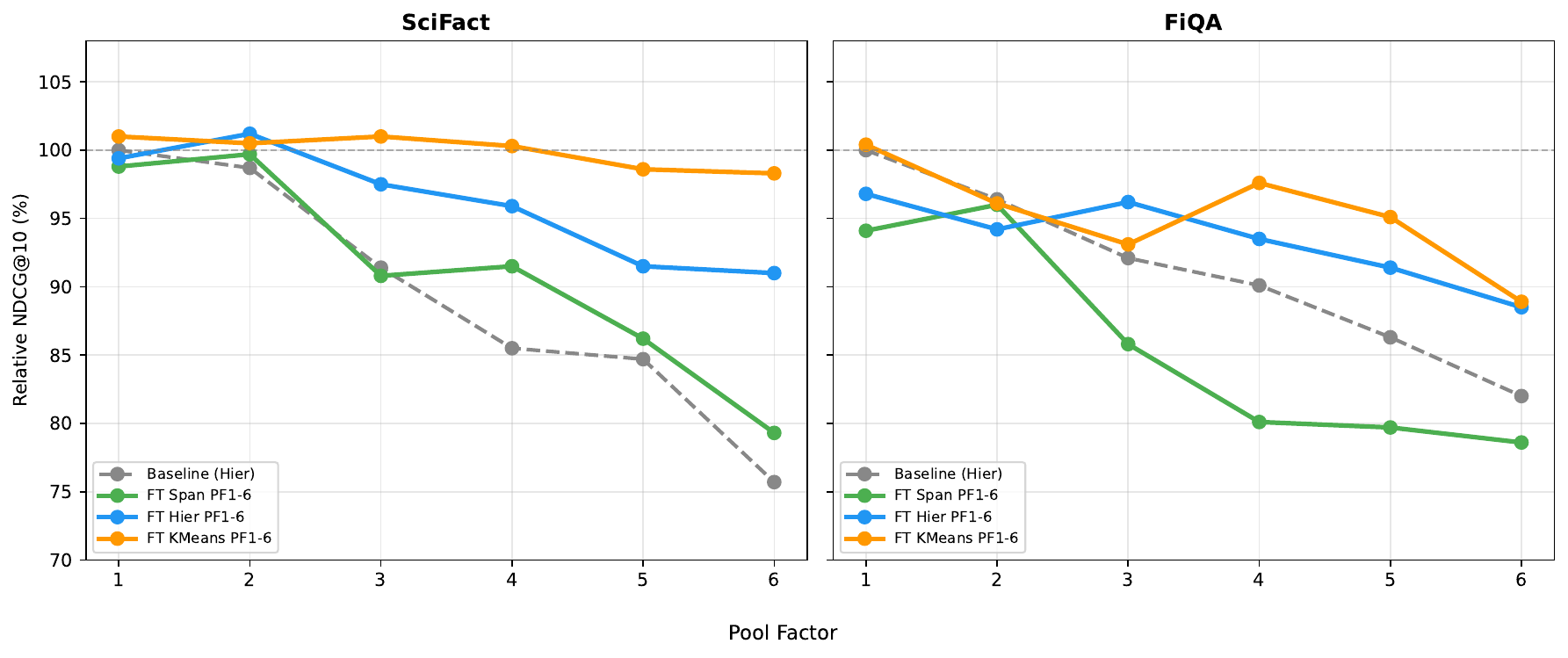}
  \caption{Cross-method comparison on NanoBEIR SciFact (left) and FiQA (right). Each model is evaluated with its own training pooling method. The baseline uses hierarchical pooling. PF = pool factor. FT = fine-tuned; PF1-6 = multi-factor training with pool factors randomly sampled per batch.}
  \label{fig:cross_method}
\end{figure}

In Figure~\ref{fig:cross_method} we compare fine-tuning on the same dataset with the different pooling methods described in Section~\ref{subsec:method_pooling}. Each method is evaluated on NanoBEIR SciFact and FiQA with the same pooling method as used during fine-tuning. To keep comparisons manageable, we only compare models fine-tuned with multi-factor training (across pool factors 1--6, randomly sampled each batch). We choose the best inference-only method, hierarchical pooling, as baseline to compare to. 

On SciFact, models fine-tuned with both hierarchical and k-means perform well compared to the baseline across all pool factors. Especially the model fine-tuned with k-means (FT KMeans PF1-6) achieves strong results, outperforming the unpooled baseline across pool factors 1--4. This implies no cost in retrieval accuracy at a vector reduction rate of 75\%, and still retaining 98.3\% relative performance at pool factor 6. 

On FiQA, the results are not as pronounced as for SciFact. The strongest model (FT KMeans PF1-6) slightly outperforms the baseline model on pool factors 1 and 3, starting to show clear gains only from pool factor 4 onwards (97.6\% vs 90.1\%). While the relative performance of 95.1\% at pool factor 5 is still a positive result, it is clearly behind the strong performance on SciFact. This shows that the effects of pooling-aware fine-tuning are dataset-dependent: SciFact documents might contain a larger share of potentially redundant tokens, while FiQA is harder to compress without incurring losses in retrieval accuracy.

When compared to a stronger baseline than sequential (span) pooling, the model fine-tuned with span pooling (FT Span PF1-6) only outperforms hierarchical inference-only pooling on SciFact for pool factors 4--6. On FiQA it even underperforms the baseline across all pool factors. While the relative gains from pooling-aware fine-tuning as described in Section~\ref{subsubsec:training_aware_results} were substantial for span pooling, it clearly lags behind the clustering-based methods on both datasets.

Two interesting points are implied by these results: 
\begin{itemize}
\item Despite trailing behind hierarchical pooling when considering inference-only regimes, k-means performs best for pooling-aware fine-tuning. Investigating why k-means is better suited for pooling-aware fine-tuning would be highly relevant, but is beyond the scope of this paper. 
\item Despite performing worst when considering pure inference pooling, through lightweight pooling-aware fine-tuning with only 809 training queries, we were able to obtain a model that achieves 75\% vector compression on SciFact at no cost to retrieval accuracy.
\end{itemize}

\subsubsection{Cross-Method Transfer}

In Table~\ref{tab:cross_method_span} we evaluate models trained with different pooling methods on a single inference pooling method. All models are evaluated using span pooling. The results show that learned pooling behavior transfers across methods. Models fine-tuned with hierarchical and k-means pooling still show large and consistent gains over the baseline on NanoBEIR SciFact. This suggests that any of the three pooling-aware fine-tuning techniques makes the token vectors more pooling-friendly overall. Fine-tuning with span pooling still shows the best results across all pooled settings (PF$\,\geq$\,2), in line with the expectation that using the same pooling method for training and inference is beneficial.

\begin{table*}
  \caption{Cross-method transfer to span pooling on NanoBEIR SciFact. All models are evaluated with span pooling, regardless of training method. FT Hier PF1-6 and FT KMeans PF1-6 were trained with hierarchical and k-means pooling, respectively. PF = pool factor. FT = fine-tuned; PF1-6 = multi-factor training with pool factors randomly sampled per batch. NDCG@10 reports absolute scores; Rel\% is relative to the baseline at PF1 (NDCG@10 = 0.808). Best NDCG@10 per pool factor in \textbf{bold}.}
  \label{tab:cross_method_span}
  \begin{tabular}{ccccccccc}
    \toprule
    & \multicolumn{2}{c}{Baseline} & \multicolumn{2}{c}{FT Span PF1-6} & \multicolumn{2}{c}{FT Hier PF1-6} & \multicolumn{2}{c}{FT KMeans PF1-6} \\
    PF & NDCG & Rel\% & NDCG & Rel\% & NDCG & Rel\% & NDCG & Rel\% \\
    \midrule
    1 & .808 & 100.0 & .799 & 98.8 & .803 & 99.4 & \textbf{.817} & 101.0 \\
    2 & .680 & 84.1 & \textbf{.806} & 99.7 & .785 & 97.1 & .801 & 99.1 \\
    3 & .526 & 65.1 & \textbf{.734} & 90.8 & .684 & 84.6 & .725 & 89.7 \\
    4 & .278 & 34.4 & \textbf{.740} & 91.5 & .565 & 70.0 & .722 & 89.3 \\
    5 & .218 & 26.9 & \textbf{.697} & 86.2 & .391 & 48.4 & .591 & 73.1 \\
    6 & .220 & 27.3 & \textbf{.641} & 79.3 & .334 & 41.4 & .555 & 68.6 \\
    \bottomrule
  \end{tabular}
\end{table*}

\subsubsection{Cross-Dataset Effects}
\label{subsubsec:cross_data_effects}

An important question with in-domain fine-tuning concerns its potential negative effects on a model's generalization capability. This is especially important for ColBERT models, which are known for their improved generalization behavior compared to single vector models. This study naturally raises the question whether continued in-domain pooling-aware fine-tuning harms performance on other datasets. 

Table~\ref{tab:cross_dataset} shows results from evaluating models fine-tuned with k-means multi-factor training (pool factors 1--6) on SciFact and FiQA across multiple NanoBEIR datasets. We observe a small negative effect on most datasets when considering unpooled performance (with the exception of in-domain evaluation on SciFact and FiQA). For the model fine-tuned on SciFact (FT SciFact KMeans PF1-6), unpooled performance on other datasets remains above 99\% relative performance compared to the baseline. For the model fine-tuned on FiQA (FT FiQA KMeans PF1-6), unpooled performance drops below 99\% for NFCorpus (98.4\%) and SCIDOCS (96.3\%).

When considering pool factors of at least 2, however, \textbf{we only observe positive effects from in-domain pooling-aware fine-tuning on other datasets}. This suggests that rather than hurting generalization, pooling-aware fine-tuning actually benefits other datasets. Instead of simply overfitting to a single dataset, the model learns to form token vectors which are generally useful for inference pooling.

The observed gains generally increase with higher pool factors. At pool factor 6, the model fine-tuned only on SciFact improves relative performance on NFCorpus from 82\% to 96.9\% and SCIDOCS from 83.8\% to 95.8\%. The model trained on FiQA lifts performance on NFCorpus at pool factor 4 from 91.4\% to 101.7\%. This implies that the pooled behavior learned by fine-tuning only on FiQA transfers remarkably well to another dataset, outperforming the unpooled baseline while achieving a vector compression rate of 75\%.

These results demonstrate that continued in-domain pooling-aware fine-tuning does not hurt, but in fact improves generalization for pool factors of 2 and above. 

\begin{table*}
  \caption{Cross-dataset effects of pooling-aware fine-tuning on NanoBEIR. All models are evaluated with k-means pooling. Each trained model was fine-tuned with k-means pooling on a single dataset (SciFact or FiQA) and evaluated on all five datasets. PF = pool factor. FT = fine-tuned; PF1-6 = multi-factor training with pool factors randomly sampled per batch. NDCG@10 reports absolute scores; Rel\% is relative to the baseline at PF1 per dataset. $\Delta$ = absolute NDCG@10 difference vs the baseline at the same PF. Best NDCG@10 per dataset and pool factor in \textbf{bold}.}
  \label{tab:cross_dataset}
  \begin{tabular}{clcccccccc}
    \toprule
    & & \multicolumn{2}{c}{Baseline (KMeans)} & \multicolumn{3}{c}{FT SciFact KMeans PF1-6} & \multicolumn{3}{c}{FT FiQA KMeans PF1-6} \\
    Dataset & PF & NDCG & Rel\% & NDCG & Rel\% & $\Delta$ & NDCG & Rel\% & $\Delta$ \\
    \midrule
    SciFact & 1 & .808 & 100.0 & \textbf{.817} & 101.0 & +.008 & .802 & 99.3 & $-$.006 \\
    & 2 & .765 & 94.7 & \textbf{.813} & 100.5 & +.047 & .774 & 95.8 & +.009 \\
    & 4 & .649 & 80.2 & \textbf{.810} & 100.3 & +.162 & .808 & 99.9 & +.159 \\
    & 6 & .609 & 75.4 & \textbf{.795} & 98.3 & +.185 & .758 & 93.8 & +.149 \\
    \midrule
    FiQA & 1 & .526 & 100.0 & .523 & 99.6 & $-$.002 & \textbf{.528} & 100.4 & +.002 \\
    & 2 & .488 & 92.9 & \textbf{.519} & 98.8 & +.031 & .505 & 96.1 & +.016 \\
    & 4 & .470 & 89.4 & .490 & 93.3 & +.021 & \textbf{.513} & 97.6 & +.043 \\
    & 6 & .431 & 82.0 & .459 & 87.4 & +.028 & \textbf{.467} & 88.9 & +.036 \\
    \midrule
    NFCorpus & 1 & \textbf{.375} & 100.0 & .372 & 99.4 & $-$.002 & .369 & 98.4 & $-$.006 \\
    & 2 & .370 & 98.8 & .370 & 98.9 & +.000 & \textbf{.378} & 101.0 & +.008 \\
    & 4 & .342 & 91.4 & .363 & 96.8 & +.020 & \textbf{.381} & 101.7 & +.039 \\
    & 6 & .307 & 82.0 & .363 & 96.9 & +.056 & \textbf{.369} & 98.4 & +.062 \\
    \midrule
    SCIDOCS & 1 & \textbf{.396} & 100.0 & .393 & 99.2 & $-$.003 & .382 & 96.3 & $-$.015 \\
    & 2 & .371 & 93.7 & \textbf{.385} & 97.2 & +.014 & .375 & 94.5 & +.003 \\
    & 4 & .347 & 87.7 & .374 & 94.5 & +.027 & \textbf{.387} & 97.7 & +.039 \\
    & 6 & .332 & 83.8 & \textbf{.380} & 95.8 & +.048 & .371 & 93.7 & +.039 \\
    \midrule
    Touché & 1 & \textbf{.596} & 100.0 & .595 & 99.8 & $-$.001 & .592 & 99.4 & $-$.004 \\
    & 2 & .597 & 100.2 & \textbf{.602} & 101.0 & +.005 & .601 & 100.8 & +.004 \\
    & 4 & .565 & 94.8 & \textbf{.594} & 99.7 & +.029 & .572 & 95.9 & +.007 \\
    & 6 & .545 & 91.5 & \textbf{.573} & 96.2 & +.028 & .571 & 95.9 & +.026 \\
    \bottomrule
  \end{tabular}
\end{table*}

\subsubsection{BEIR Validation}
\label{subsubsec:beir_validation}

In this section we aim to validate the main findings from the NanoBEIR results above on full BEIR evaluation runs over four datasets: FiQA, NFCorpus, SCIDOCS and SciFact. We compare 3 untrained inference-only baselines as described in Section~\ref{subsec:method_pooling}: sequential (span), hierarchical and k-means pooling. Moreover, we compare all trained multi-factor models.

When considering inference-only pooling methods, we conclude that the ranking observed on NanoBEIR holds up: hierarchical pooling is generally the best inference pooling method, closely followed by k-means, while span pooling performs significantly worse, with performance dropping off much faster than clustering-based methods on higher pool factors. 

What clearly differs is the relative performance of inference-only pooling. Full BEIR datasets appear to be much more pooling-friendly than their NanoBEIR counterparts. On NFCorpus and SciFact, hierarchical pooling remains above 95\% of unpooled performance through pool factors 1--6. These results are closer to the strong results reported by \citet{clavie2024pooling}. On NanoBEIR, hierarchical pooling falls below 90\% at pool factors 3 (NFCorpus) and 4 (SciFact), ending up much lower at pool factor 6, with 84.1\% and 75.7\%, respectively. Similar effects, though smaller, hold for FiQA and SCIDOCS.

When evaluating the results from pooling-aware fine-tuning on BEIR, SciFact confirms the positive effects from pooling-aware fine-tuning observed on NanoBEIR, while on FiQA the results are mixed.

On SciFact, the model fine-tuned with k-means outperforms the unpooled baseline across all pool factors. This suggests that we can achieve an 83\% reduction in vector count with no degradation in retrieval accuracy. Fine-tuning with hierarchical pooling achieves similarly strong performance, though underperforming k-means at pool factors 1, 5 and 6 and falling slightly below the unpooled baseline at pool factor 6 (99.3\%). Fine-tuning with span pooling delivers strong gains over inference-only span pooling but remains behind stronger inference-only methods (hierarchical, k-means) at pool factors of 3 and above. 

On FiQA, no fine-tuned multi-factor model outperforms strong inference-only methods (hierarchical, k-means) at pool factor 2. From pool factor 3 onwards, we see clear improvements in the model trained with k-means, which retains more than 97\% of unpooled performance until pool factor 4, then drops to 93.8\% and 88.9\% for pool factors 5 and 6, respectively. However, we don't observe improvements from fine-tuning with hierarchical or span pooling at low pool factors -- where they actually perform worse than their training-free baselines -- with gains emerging only at pool factors 4--6 for hierarchical and pool factors 4 and 6 for span pooling. 

When contrasting these results with NanoBEIR, we conclude that for SciFact, the results on BEIR are even better than on NanoBEIR. This is intuitive, since inference-only pooling methods also perform better on BEIR SciFact than on NanoBEIR SciFact. On FiQA on the other hand, the models trained with k-means and hierarchical pooling perform roughly similarly across BEIR and NanoBEIR datasets, but since inference-only pooling performs better on BEIR FiQA, the relative gains from pooling-aware fine-tuning are smaller, though still substantial for k-means. 

This further suggests that the positive effects from pooling-aware fine-tuning are dataset-dependent and that fine-tuning with k-means delivers gains in a consistent way, while we cannot conclude the same from fine-tuning with hierarchical and span pooling. 

Finally, we confirm that the positive generalization effect across datasets observed in Section~\ref{subsubsec:cross_data_effects} holds true in BEIR. Training on one dataset with k-means (SciFact, FiQA) generally improves pooled performance on other datasets. Both models show a positive transfer to each other's in-domain datasets. When evaluated on NFCorpus, the model fine-tuned on SciFact outperforms the unpooled baseline on pool factors 1--3, then drops slightly below 100\% and remains strong at pool factor 6 with a relative performance of 97.6\%. The model fine-tuned on FiQA performs similarly strong on NFCorpus. While both models do not beat the strongest inference pooling method (hierarchical) on SCIDOCS (except for pool factor 2), they still outperform the untrained k-means baseline on pool factors 2--5.

\begin{figure}
  \centering
  \includegraphics[width=\linewidth, alt={Four line charts showing BEIR validation results on FiQA, NFCorpus, SCIDOCS, and SciFact. Dashed lines show three inference-only pooling baselines and solid lines show fine-tuned multi-factor models, which consistently outperform inference-only methods on SciFact and show positive transfer to other datasets across pooled settings.}]{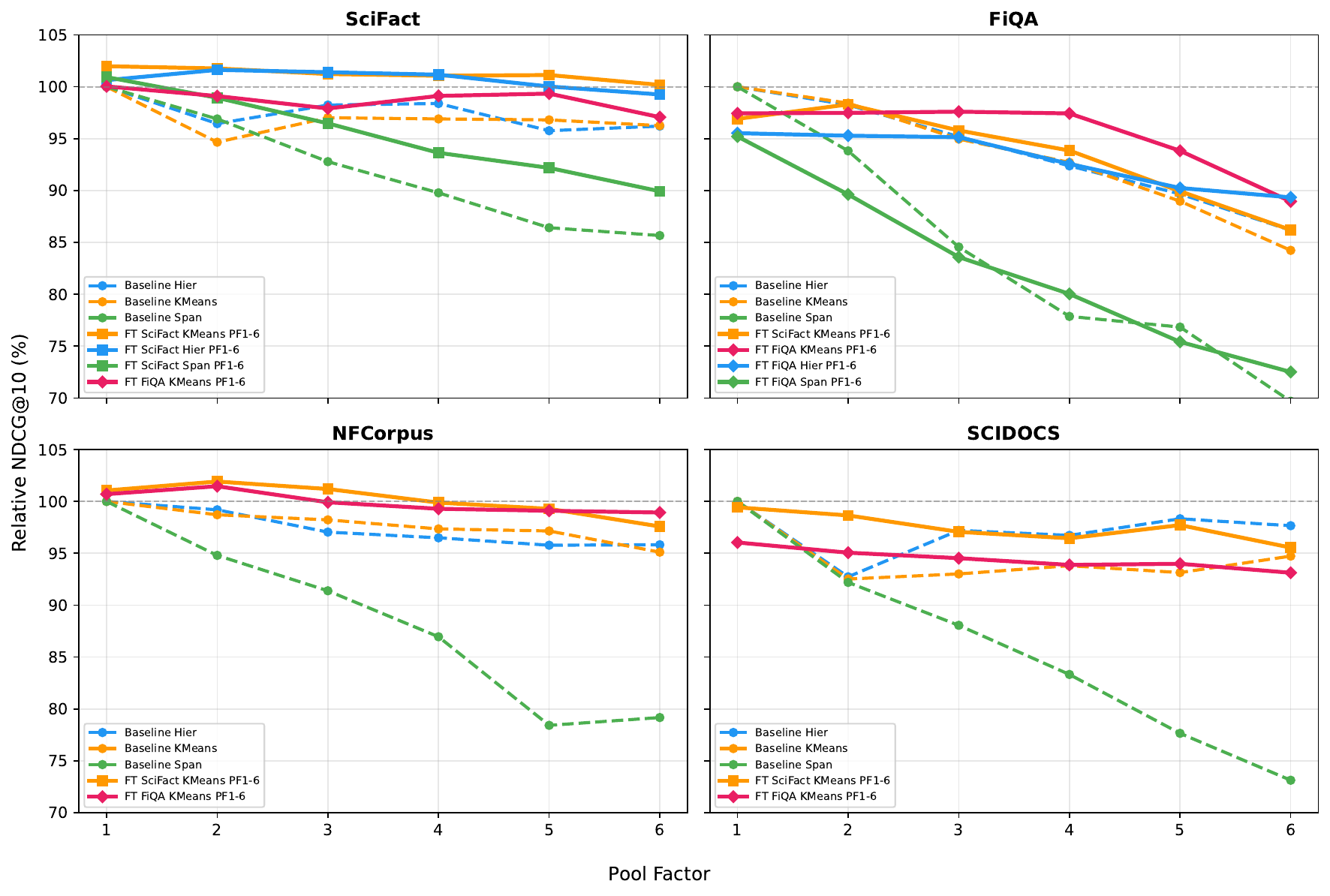}
  \caption{BEIR validation of inference-only pooling and pooling-aware fine-tuning. Each panel shows one BEIR dataset. Dashed lines: baseline model with three inference-only pooling methods (hierarchical, k-means, span). Solid lines: SciFact- and FiQA-trained PF1-6 models evaluated with their respective pooling method. PF = pool factor. FT = fine-tuned; PF1-6 = multi-factor training with pool factors randomly sampled per batch.}
  \label{fig:beir_validation}
\end{figure}

\section{Limitations}

This work was carried out as an independent research project with limited compute budget.\footnote{All training and evaluation runs were conducted on a single RTX A4500 GPU.} Thus, during model training and evaluation we focused on small models and datasets. We fine-tuned a small ColBERT model on two small BEIR datasets (SciFact and FiQA) and evaluated the resulting models mainly on selected NanoBEIR datasets, while validating the main findings on selected BEIR datasets. All results are from single training runs, except the SciFact multi-factor models, which report averaged results across two training runs. 

To increase and test the robustness of the results, it would be beneficial to run a larger number of experiments using more and larger datasets, full BEIR evaluation runs, different models, fine-tuned across various pool factors, with multiple averaged runs per setting. Moreover, since this work focuses exclusively on fine-tuning existing ColBERT models, a promising direction would be to integrate multi-factor pooling-aware training into full ColBERT pre-training, which could yield a general-purpose model where the desired compression level is chosen at indexing time without any additional fine-tuning.

Pooling-aware fine-tuning adds non-trivial training overhead: for multi-factor training, span pooling increases training time by roughly $3\times$ relative to training without pooling, hierarchical pooling by approximately $7$--$9\times$, and k-means pooling by $10$--$13\times$. However, this additional training cost can be a worthwhile trade-off, as it reduces document storage requirements and scoring cost at inference time.

\section{Conclusion}

To address the high storage and memory costs of late interaction models, in this paper we investigated whether lightweight pooling-aware fine-tuning can lead to improvements over inference-only pooling. While hierarchical pooling is the strongest inference-only method, k-means performs best for pooling-aware fine-tuning, producing robust gains across datasets and pool factors. Results for hierarchical and span fine-tuning are more dataset-dependent: hierarchical fine-tuning can be very strong in some settings, but offers limited improvement in others. Span pooling shows the largest relative gains on NanoBEIR, but does not hold up consistently on BEIR and remains the weakest method overall.

We present evidence that training on one pool factor improves performance across other pool factors and that multi-factor training is an effective technique to obtain a model that works across multiple pool factors during inference. Further, we observe a positive transfer when training and evaluation use different pooling methods, and show that pooling-aware fine-tuning with k-means generalizes across datasets, improving pooled performance on datasets absent from continued fine-tuning.

One direction for future research we are particularly excited about is the idea of dynamic vector allocation. Rather than fixing or randomly sampling a pool factor during training, it would be beneficial to choose the right pool factor for a given document based on its complexity and information-density. 

\begin{acknowledgments}
We thank Answer.AI, LightOn, Mixedbread, PyLate and Sentence Transformers for their open-source contributions that made this research possible.
\end{acknowledgments}

\section*{Declaration on Generative AI}
 During the preparation of this work, the author used Claude 4.6 Opus and SolveIt in order to: Grammar and spelling check. Further, the author used Claude 4.6 Opus and SolveIt for figures 1--4 and tables 1--4 in order to: Formatting assistance. After using these tools, the author reviewed and edited the content as needed and takes full responsibility for the publication’s content. 

\bibliography{references}

\clearpage
\appendix

\section{Additional Results}
\begin{table}[ht]
  \caption{Effect of pooling-aware fine-tuning with hierarchical pooling on NanoBEIR SciFact. All models are evaluated with hierarchical pooling. PF = pool factor. FT = fine-tuned; FT PF1 = fine-tuned without pooling during training; PF1-6 = multi-factor training with pool factors randomly sampled per batch. NDCG@10 reports absolute scores; Rel\% is relative to the baseline at PF1 (NDCG@10 = 0.808). Best NDCG@10 per pool factor in \textbf{bold}.}
  \label{tab:hier_training}
  \begin{tabular}{ccccccccc}
    \toprule
    & \multicolumn{2}{c}{Baseline} & \multicolumn{2}{c}{FT PF1} & \multicolumn{2}{c}{FT Hier PF2} & \multicolumn{2}{c}{FT Hier PF1-6} \\
    PF & NDCG & Rel\% & NDCG & Rel\% & NDCG & Rel\% & NDCG & Rel\% \\
    \midrule
    1 & .808 & 100.0 & \textbf{.819} & 101.3 & \textbf{.819} & 101.3 & .803 & 99.4 \\
    2 & .798 & 98.7 & .772 & 95.5 & \textbf{.838} & 103.7 & .818 & 101.2 \\
    3 & .739 & 91.4 & .286 & 35.3 & \textbf{.804} & 99.5 & .788 & 97.5 \\
    4 & .691 & 85.5 & .113 & 14.0 & .742 & 91.9 & \textbf{.775} & 95.9 \\
    5 & .685 & 84.7 & .241 & 29.8 & .711 & 88.0 & \textbf{.739} & 91.5 \\
    6 & .612 & 75.7 & .330 & 40.9 & .718 & 88.9 & \textbf{.735} & 91.0 \\
    \bottomrule
  \end{tabular}
\end{table}

\begin{table}[ht]
  \caption{Effect of pooling-aware fine-tuning with k-means pooling on NanoBEIR SciFact. All models are evaluated with k-means pooling. PF = pool factor. FT = fine-tuned; FT PF1 = fine-tuned without pooling during training; PF1-6 = multi-factor training with pool factors randomly sampled per batch. NDCG@10 reports absolute scores; Rel\% is relative to the baseline at PF1 (NDCG@10 = 0.808). Best NDCG@10 per pool factor in \textbf{bold}.}
  \label{tab:kmeans_training}
  \begin{tabular}{ccccccccc}
    \toprule
    & \multicolumn{2}{c}{Baseline} & \multicolumn{2}{c}{FT PF1} & \multicolumn{2}{c}{FT KMeans PF2} & \multicolumn{2}{c}{FT KMeans PF1-6} \\
    PF & NDCG & Rel\% & NDCG & Rel\% & NDCG & Rel\% & NDCG & Rel\% \\
    \midrule
    1 & .808 & 100.0 & \textbf{.819} & 101.3 & .806 & 99.7 & .817 & 101.0 \\
    2 & .765 & 94.7 & .676 & 83.6 & \textbf{.815} & 100.8 & .813 & 100.5 \\
    3 & .756 & 93.6 & .260 & 32.2 & \textbf{.817} & 101.0 & .816 & 101.0 \\
    4 & .649 & 80.2 & .102 & 12.7 & .797 & 98.7 & \textbf{.810} & 100.3 \\
    5 & .645 & 79.8 & .252 & 31.2 & .777 & 96.1 & \textbf{.797} & 98.6 \\
    6 & .609 & 75.4 & .292 & 36.1 & .722 & 89.4 & \textbf{.795} & 98.3 \\
    \bottomrule
  \end{tabular}
\end{table}
\end{document}